# Dynamical Casimir effect for surface plasmon polaritons

V. Hizhnyakov, A. Loot, S.Ch. Azizabadi

Institute of Physics, University of Tartu, Ravila 14c, 50411 Tartu, Estonia
E-mail: hizh@fi.tartu.ee

**Abstract.** Emission of photon pairs by an interface of asymmetric dielectric and thin metal film excited by a normally falling plane wave is considered. The excitation causes oscillations in time of the phase velocity of surface plasmon polaritons in the interface. This leads to the dynamical Casimir effect – the generation of pairs of surface plasmon polariton quanta, which transfer to photons outside the interface. In case of a properly chosen interface, the yield of two-photon emission may exceed that of usual spontaneous parametric down conversion.

## 1. Introduction

In this communication we consider emission of photon pairs by a metal-dielectric interface exposed into strong laser wave with the wave front parallel to the interface. The laser wave, due to the nonlinear interaction of light with matter, evokes periodical oscillations in time of the optical length of the electromagnetic excitations of the interface - surface plasmon polaritons (SPPs). The latter changes cause periodical perturbation in time of the zero-point state of the excitations resulting in the emission of pairs of SPP quanta. This emission is fully analogous to the dynamical Casimir effect (DCE) - generation of pairs of photons in a resonator when changing its optical length in time. (see, e.g. [1 - 10]).

To observe DCE, one needs to move the mirror(s) of the resonator with velocity comparable to velocity of light [1]. This condition is difficult to fulfill. Therefore the intensity of corresponding emission is usually extremely weak. Nevertheless by means of a superconducting circuit consisting of a coplanar transmission line one can achieve for microwaves rather strong change of the electrical length in time. Recently in this way the DCE was observed in 10 GHz diapason [11]. In the visible the DCE may be also enhanced if to use strong laser excitation. However the electric field $E$ associated with visible photons is very weak. Therefore one needs to use very strong laser excitation to observe the DCE [12-14].

A possibility to overcome this difficulty is given by plasmonics: the electromagnetic excitations of the metal-dielectric interface, SPPs, have much stronger electric field compared to a photons in free space or in usual dielectrics [15, 16]. This allows one to expect that the yield of the DCE-induced emission in metal-dielectric interface may be essentially increased. The goal of this communication is to consider this possibility. We will show that the electric field enhancement of SPPs can indeed lead to strong enhancement of the yield of the DCE.

## 2. Scheme of proposed experiment

We consider a metal-dielectric interface (refractive indices $n_m$ and $n_1$ respectively) in the $(x, y)$ plane with the length $l_1$ (along y-axis) and width $l_0$ (see Fig. 1a). It is supposed that the electric susceptibility of the dielectric is diagonal in the reference frame under consideration and the dielectric has no inverse symmetry in $(x, y)$ plain. In this case SPPs in the interface have $z$ polarization and the second-order nonlinear susceptibility in the interface $\chi^{(2)}_{\alpha,zz}$ differs from zero. The interface is irradiated by a monochromatic laser wave with frequency $\omega_0$ (or wavelength $\lambda_0$) and $y$ polarization falling normally to the interface falling normally to the interface (the wave front of the excitation is parallel to the interface). The structure may include additional dielectric layer(s) between the metallic film and the prism, see Fig. 1b for example, in order to increase the propagation length of SPPs $\Delta l = \lambda/(2\pi \operatorname{Im} n)$, where $n$ is the effective mode index of SPPs [17-19].

To be specific, we note that in the case of silver-air interface the propagation length of SPPs at $\lambda = 1000$ nm equals 1.5 mm (see Fig. 2a). However it may be more than one order of magnitude larger for long-range SPPs if additional dielectric film is added (see Fig. 2b). The data presented in Fig. 2a and 2b are calculated by analytic dispersion relation formula for SPPs for 2 and 3 layer structure [15]. The refractive index for silver is taken from [20]. Note that the real part of the effective mode index of SPPs and long-range SPPs is close to unity; see Figs. 2a and 2b.

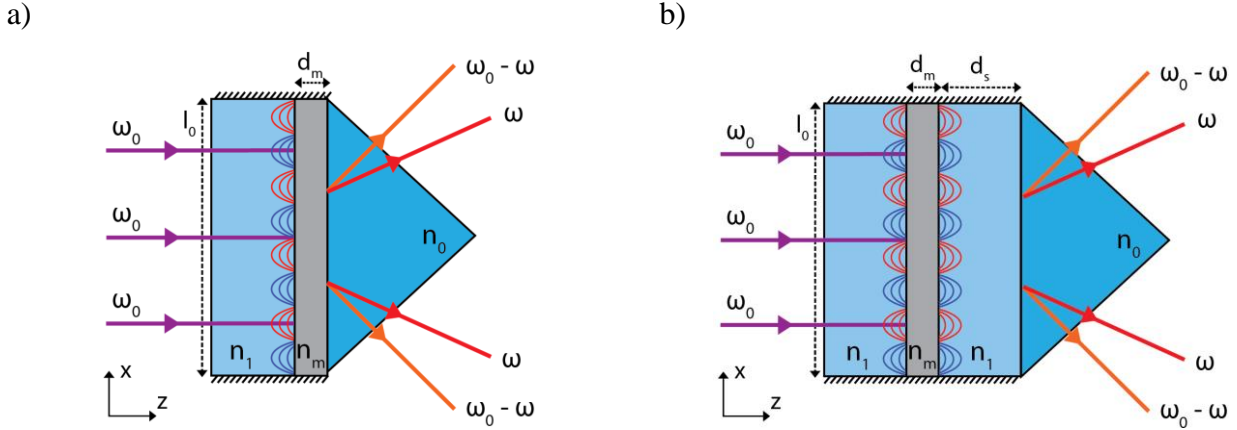

Fig. 1. Two-photon emission due to the DCE in the dielectric-metal interface. The scheme consists of prism, thin metal film and dielectric which refractive indices are denoted by $n_0$, $n_m$ and $n_1$ respectively (Fig. 1a). The thickness of metal film is $d_m$ and the length of the metal-dielectric interface is $l_0$. The interface is placed in a resonator and evenly illuminated by laser light with frequency ω₀ at normal angle of incidence. The generated SPPs (with frequencies $\omega_0 - \omega$ and $\omega$) can leave the interface through the prism as photons in the direction corresponding to the Kretschmann configuration. In Fig. 1b additional layer with thickness $d_s$ is added between the prism and the metal film to excite long-range SPPs.

In the case under consideration the second-order nonlinear susceptibility $\chi^{(2)} \equiv \chi^{(2)}_{y,zz}$ in the metal-dielectric interface differs from zero. Therefore the excitation with $y$ polarization causes the time-dependence of the effective mode index of SPPs with $z$ polarization. The time dependent part of it equals

$$\delta n_t = \chi^{(2)} E_0 \cos(\omega_0 t). \qquad (1)$$

Here $E_0 = \sqrt{Z_0 I_0}$ is the amplitude of the electric field of laser wave, $Z_0 = 376.7\,\Omega$ is the impedance of free space, $I_0$ is the intensity of the laser light, $\omega_0$ is its frequency and $\chi^{(2)}$ is the second order nonlinear susceptibility (it is assumed that $|\delta n_t| \ll 1$).

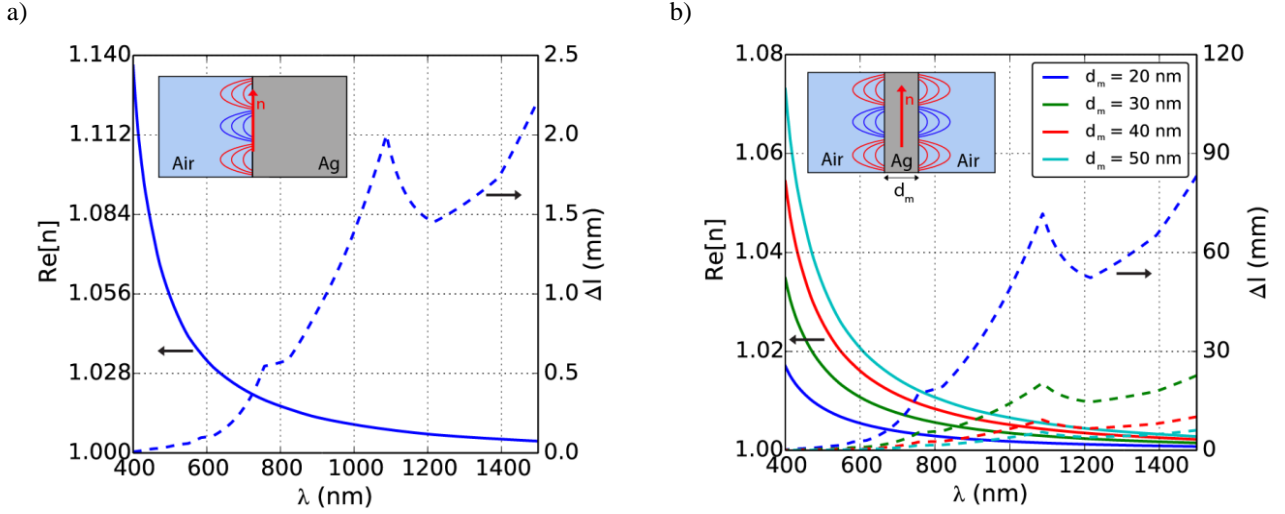

Fig. 2. The effective mode index real part $\mathrm{Re}[n]$ (solid lines) and propagation distance $\Delta l$ (dashed lines) of SPPs at air-silver interface (a) and of long-range SPPs at air-silver-air structure (b). The $d_m$ denotes the thickness of the silver film and $\lambda$ is the wavelength of light. The structure under investigation is also presented in the inset.

Consequently, due to nonlinear interaction of light with matter the laser wave induces periodical oscillations in the effective mode index of SPPs in time, being the same for all coordinates of the interface. These oscillations induce a periodic variation of the optical length of SPPs in the resonator. According to the DCE these oscillations cause generation of pairs of quanta of SPP. The generated SPPs can leave the interface through the dielectric prism as photons in direction $\varphi_0 = \arccos(n/n_0)$ corresponding to the Kretschmann configuration, as it is depicted on Fig. 1 (in this direction the tangential wave vector is conserved).

We suppose that one of the mirror in Fig. 1 is situated at coordinate $x = 0$ and another at $x = l_0$. Then, due to the oscillations of the effective mode index the optical length of SPPs $L_t$ changes periodically in time:

$$L_t = L + a\cos(\omega_0 t). \qquad (2)$$

Here $L = l_0 n \approx l_0$ where $n$ is the effective mode index of the SPPs under consideration,

$$a = l_0 |\chi^{(2)}| E_0 \qquad (3)$$

is the amplitude of oscillations in the optical length. Here we neglect small dependence of the optical length on the frequency of the SPPs under consideration. We also neglect the damping of the SPPs supposing that $l_0$ is less than the propagation distance $\Delta l$ of the SPPs (see Fig. 2).

### 3. Equation for field operator

Let us consider the generation of SPPs with the wave vector $\vec{\kappa}$ with components $\kappa_x = \pi k/L$ and $\kappa_y = \pi q/l_1$, where $k, q = 0, 1, 2, ...$. To find the number of generated quanta of a mode $(k, q)$ due to the DCE one should calculate the Bogoliubov transformation of the field operators $\hat{b}^{(1)}_{k,q} = \mu_{k,q} \hat{b}_{k,q} + \nu_{k,q} \hat{b}^+_{k,q}$ [11, 12]. Here $\hat{b}_{k,q}$ and $\hat{b}^{(1)}_{k,q}$ are the annihilation operators of the mode $(k, q)$ at small and large time, respectively. $\mu_{k,q}$ and $\nu_{k,q}$ are the complex parameters satisfying the condition $|\mu_{k,q}|^2 = 1 + |\nu_{k,q}|^2$. The number of generated photons is $N_{k,q} = |\nu_{k,q}|^2$. Another, although similar way to find $N_{k,q}$ is to calculate the large time asymptotic of the correlation function $d_{k,q}(\tau)$ of the field operator of a mode $\hat{A}_{k,q}(t) = \sqrt{\hbar/2\omega_{k,q}} \hat{b}^{(1)}_{k,q} e^{-i\omega_{k,q}t}$ [21]. In the large $t$ limit this function equals to

$$d_{k,q}(\tau) = \Theta(\tau)\langle 0|\hat{A}^+_{k,q}(t+\tau)\hat{A}_{k,q}(t)|0\rangle = (\hbar/2\omega_{k,q}) N_{k,q} e^{i\omega_{k,q}\tau}, \quad t \to \infty, \tag{4}$$

where $|0\rangle$ is the initial zero-point state. The field operator $\hat{A}$ of SPPs satisfies the wave equation

$$\ddot{\hat{A}} - c^2 \left(\partial^2/\partial x^2 + \partial^2/\partial y^2\right)\hat{A} = 0 \tag{5}$$

and the boundary conditions $\hat{A}(x=0, t) = 0$ and $\hat{A}(x=L_t, t) = 0$. Here $x, y$ are the coordinates at the interface, $c$ is the phase velocity of the SPP under consideration (the weak dependence of $c = c_0/n$ of SPPs of frequency is neglected). Therefore this operator can be presented as the linear combination of the plane wave operators. Here we take

$$\hat{A}(x, y, t) = \sum_{k,q} \hat{A}_{k,q} e^{i\kappa_y y} \sin(\pi k x/L_t), \tag{6}$$

where $\hat{A}_{k,q}$ is the field operator of the mode $(k, q)$. Inserting this operator to the wave equation we get in the large $L/\lambda$ limit

$$\sum_{k=-\infty}^{\infty}\left[\left(\ddot{\hat{A}}_{k,q} + \omega^2_{k,q}\hat{A}_{k,q}\right)\sin(\pi k x L_t^{-1}) - k\pi x L_t^{-2}\left(2\dot{L}_t\dot{\hat{A}}_{k,q} + \ddot{L}_t\hat{A}_{k,q}\right)\cos(\pi k x L_t^{-1})\right] = 0 \tag{7}$$

Here

$$\omega_{k,q} = c\pi\sqrt{(k/L)^2 + (q/l_1)^2} \tag{8}$$

is the frequency of the mode $(k,q)$ (the terms $\propto L_t^{-m}$ with $m \gg 3$ are neglected, these terms in case $L/\lambda \gg 1$ are small). Now we use the equation for the Fourier series of the saw-tooth wave

$$x = -\sum_{j \neq 0}(-1)^j j^{-1} \sin(jx), \quad x - \pi \in 2\pi n$$

($n = \pm 1, \pm 2,...$), which gives for $\hat{A}_{k,q} = (-1)^k \hat{A}_{k,q}$ the equation

$$\ddot{\hat{A}}_{k,q} + \omega_{k,q}^2 \hat{A}_{k,q} = \omega_{k,q} \hat{B}_{k,q}, \qquad (9)$$

where

$$\hat{B}_{k,q} = \frac{1}{\pi c} \sum_{j \neq k} \frac{j(2\dot{L}_t \dot{\hat{A}}_{j,q} + \ddot{L}_t \hat{A}_{j,q})}{j^2 - k^2}. \qquad (10)$$

The operator $\hat{B}_{k,q}$ describes the effect of oscillations in the optical length. In the integral form this equation reads as

$$\hat{A}_{k,q}(t) = \hat{A}_{k,q}^{(0)}(t) - \int_{-\infty}^{t} \sin(\omega_{k,q}(t - t')) \hat{B}_{k,q}(t') dt', \qquad (11)$$

where $\hat{A}_{k,q}^{(0)}(t)$ is the operator of the undisturbed field.

### 4. Generation rate of SPP quanta

In the $t \to \infty$ limit the nonzero contribution to the integral in Eq. (11) comes from the terms $\propto e^{\pm i(\omega_0 - \omega_{k,q} - \omega_{j,q})t}$ with $\omega_{j,q} = \omega_0 - \omega_{k,q}$. This means that we can take

$$2\dot{L}_t \dot{\hat{A}}_{j,q} + \ddot{L}_t \hat{A}_{j,q} \cong (2\omega_0 \omega_{j,q} - \omega_0^2) a \cos(\omega_0 t) \hat{A}_{j,q} = a(\pi c/L)^2 (j^2 - k^2) \cos(\omega_0 t) \hat{A}_{j,q} \qquad (12)$$

As a result, the factor $j^2 - k^2$ in the equation for $\hat{B}_{k,\bar{q}}(t)$ cancels and we get $\hat{B}_{k,q}(t) \cong \hat{B}_q(t)$, where

$$\hat{B}_q(t) = 2a \cos(\omega_0 t) \hat{Q}_q, \quad \hat{Q}_q = L^{-1} \sum_{j=1}^{\infty} \omega_j \hat{A}_{j,q}(t). \qquad (13)$$

Here $\omega_j = \pi c j / L$. One can see that in the case of monochromatic laser excitation in the large time limit the $k$-dependence of $\hat{B}_{k,q}(t)$ disappears.

Inserting Eqs. (11) and (13) into Eq. (4) the negative frequency term $\propto e^{i\omega_{k,q}\tau}$ in the correlation function $d_k(\tau)$ comes from the $\propto \sin(\omega_k(t + \tau - t'))$ term. Therefore

$$N_{k,q}(t) \simeq \frac{a^2 \omega_{k,q}}{2\hbar L} \int_{-\infty}^{t} dt_1 \int_{-\infty}^{t} dt_1' e^{i(\omega_0 - \omega_{k,q})(t_1 - t_1')} D_q(t_1, t_1'). \qquad (14)$$

Here $\omega_k = \pi c k / L_0$,

$$D_q(t_1, t_1') = L \langle 0 | \hat{Q}_q^+(t_1) \hat{Q}_q(t_1') | 0 \rangle \qquad (15)$$

is the correlation function of the perturbation of the field. The DCE-caused emission of photon pairs results in small (as compared to the frequencies) decay of the excited modes of SPPs. This decay may be taken into account phenomenologically by adding small imaginary part to the frequencies of the modes. The spectrum of the emission under consideration is continuous with the width being comparable to $\omega_0$. Therefore the effect of small decay of the modes in our case is negligibly small.

Taking into account that in the $t_1, t_1' \to \infty$ limit the pair-correlation function depends on the time difference ($D_q(t_1, t_1') = D_q(t_1 - t_1')$) we get the following equation for the emission rate

$$\dot{N}_{k,q} = \left(a^2/2\hbar L\right)\omega_{k,q} D_q(\omega_0 - \omega_{k,q}), \tag{16}$$

where

$$D_q(\omega_0 - \omega_{k,q}) = \int_{-\infty}^{\infty} d\tau\, e^{i(\omega_0 - \omega_{k,q})\tau} D_q(\tau). \tag{17}$$

Here we consider the case of small amplitude of oscillations in the optical length as compared to the wavelength $\lambda = 2\pi/\omega$ of the SPPs, i.e. $a \ll \lambda$. In this case the velocity of oscillations in the optical length $V = a\omega$ is small as compared to the phase velocity $V \ll c$. (The case $V \sim c$ was considered in Refs. [9, 10].) Then in the equation for $D_q(t - t')$ one can take

$$Q(t) \approx Q^{(0)}(t) = L^{-1}\sum_{j=1}^{\infty} \omega_j \hat{A}_j^{(0)}(t). \tag{18}$$

This gives

$$D_q(\tau) \approx D_q^{(0)}(\tau) = \frac{\hbar}{2L}\sum_j \frac{\omega_j^2}{\omega_{j,q}} e^{-i\omega_{j,q}\tau}. \tag{19}$$

In the large $L/\lambda$ limit the sum over $j$ may be replaced by the integral over $\omega' = \pi cj/L$. Taking into account the relation $dj = L d\omega'/\pi c$ we get

$$D_q(\omega_0 - \omega_{k,q}) = \hbar c^{-1}\sqrt{(\omega_0 - \omega_{k,q})^2 - \omega_q^2}, \tag{20}$$

where $\omega_q = \pi cq/L$. As a result we get the following equation for the rate of generation of SPP pairs:

$$\dot{N}_{k,q} \approx \left(a^2/2Lc\right)\omega_{k,q}\sqrt{(\omega_0 - \omega_{k,q})^2 - \omega_q^2} \tag{21}$$

in case $V/c \ll 1$. The number of working modes equals to $\sum_{k,q} 1 = \omega_0^2 L l_1/4\pi c^2$. Consequently the rate of generation of SPP pairs is linearly dependent on the length $l_1$ of the interface in $y$ direction.

In the $L, l_1 \gg \lambda$ limit the sum $\Sigma_{k,q}$ can be replaced by the integral $\iint dk dq$. Taking into account that $dk dq = \omega L l_1 d\omega d\vartheta/\pi c^2$, where $\vartheta = \arcsin(\kappa_y/\kappa)$ is the angle between the direction of emission and the normal to the interface we get

$$\dot{N}(\omega, \vartheta) = \frac{4\pi^2 a^2 l_1}{\lambda_0^3}\left(\frac{\omega}{\omega_0}\right)^3 \sqrt{\left(1 - \frac{\omega}{\omega_0}\right)^2 - \left(\frac{\omega}{\omega_0}\right)^2 \sin^2(\vartheta)}. \tag{22}$$

It is dimensionless quantity, as it should be. The dependence of the emission of SPPs on frequency and angle is presented in Fig. 4.

The total rate of emission of pairs of SPP quanta equals to

$$\dot{N} = 2\int_0^{\vartheta_0} d\vartheta \int_0^{\omega_0/(1+\sin\vartheta)} d\omega\, \dot{N}(\omega, \vartheta) \approx 2\alpha_{\vartheta_0} a^2 l_1 \omega_0/\lambda_0^3, \tag{23}$$

where $\vartheta_0 = \arctan(l_1/L)$, $\alpha_{\vartheta_0}$ is the dimensionless parameter of the order of 1. E.g. $\alpha_{\pi/4} \approx 0.81$ while $\alpha_{\pi/2} \approx 1.05$

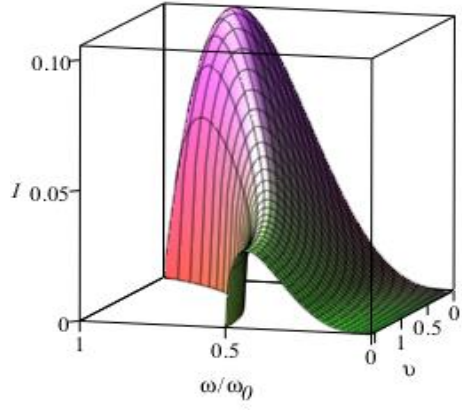

Fig. 4. Spectral and angular dependence of the DCE caused emission rate.

## 5. Where is the DCE-caused radiation generated?

From Eqs. (9) and (13) it follows that the DCE-caused radiation arises from the periodical perturbation in time of the quantum modes described by $\hat{B}_q(t)$, the for the modes with different $k$. This means that the perturbation is highly localized in the $x$ direction. Obviously this perturbation comes from the mirrors. Consequently the DCE-caused radiation source is localized at the mirrors. This means that the radiation is emitted by the mirrors and then fills up all the space between the mirrors. In current consideration the decay of modes was not considered. If the decay is taken into account then only the areas of the propagation distance at the mirrors will contribute to the oscillations of the optical length and to the DCE-caused radiation. Analogously only the emission with the angle $\vartheta \lesssim \pi/4$ corresponding to $l_1 \sim \Delta l$ will have remarkable intensity.

## 6. Yield of dynamical Casimir effect at metal-dielectric interface

Taking into account the relation (3) we get

$$\dot{N} \sim \left(l_0^2 l_1 / \lambda_0^3\right) \left|\chi^{(2)}\right|^2 I_0 Z_0. \quad (24)$$

The total number of generated pairs of quanta in the time interval $t_0$ equals to $\dot{N} t_0$. In the same time interval the number of photons falling to the excitation area $l_0 l_1$ equals to $N_0 = I_0 t_0 l_0 l_1 / \hbar \omega_0$. This gives for the yield of emission of photon pairs $\kappa = N/N_0$ the equation

$$\varsigma \sim \hbar \omega_0^2 l_0 \left|\chi^{(2)}\right|^2 Z_0 / \lambda_0^3. \quad (25)$$

The Eq. (25) is presented in the form allowing one to easily see that $\kappa$ is a dimensionless quantity: the dimensionality of $\hbar \omega_0^2$ is W, the dimensionality of $\hbar \omega_0^2 Z_0$ is $V^2$, and the dimensionality $\hbar \omega_0^2 Z_0 / \lambda_0^2$ is $(V/m)^2$, i.e. the opposite to the dimensionality of $\left|\chi^{(2)}\right|^2$.

In dielectrics the typical value of $\left|\chi^{(2)}\right|$ is 1 pm/V. In case of $\lambda_0 = 500\,nm$ for a thin and narrow dielectric plate with the excited length $l_0 = 1$ mm one gets the yield $\varsigma \sim 10^{-10}$. In reality one should expect the existence of some losses, e.g. in transformation of SPPs to photons. Therefore more realistic estimation would be $\varsigma \sim 10^{-11} \div 10^{-12}$, which is roughly the same as in case of the spontaneous parametric down conversion in dielectrics [22]. This means that for properly focused laser light with intensity ~1 microwatt one can get few pairs of photons per second. This is rather weak emission which is difficult to observe. However in the case of metal-dielectric interface the emission may be much stronger due to the strong enhancement of the electric field. The enhancement of the DCE can be described by the factor $P \sim \left(\eta_0(\omega_0)\eta^2(\omega_0/2)\right)^2$, where $\eta_0(\omega_0)$ is the renormalization of the laser field with frequency $\omega_0$ at the interface in the case of normal incidence, $\eta(\omega_0/2)$ is the enhancement of the field of the SPPs with the frequency $\omega_0/2$. An analogous equation was found in Ref. [23] for the yield of the spontaneous parametric down conversion in the metal-dielectric interface under the phase matching condition.

In this study the enhancement factor of SPPs in stratified structure was calculated by transfer-matrix method [24, 25], which is simple analytical method to solve the field distributions in layered structures. The enhancement factors (at optimal excitation of SPPs) for prism-silver-air structure for different thickness of silver film $d_m$ is presented in Fig. 4a. The refractive index of silver $n_m$ is taken form [20], the refractive index of prism is taken to be $n_0 = 1.5$ and $n_1 = 1.0$. The maximum value $\eta \approx 30$ is found in the blue spectral region for silver film thickness $d_m = 60\,nm$.

In the case of structure in Fig. 1a the propagation length of SPPs at $\lambda \approx 1000$ nm equals to 1.5 mm (see Fig. 2a). This allows one to estimate the yield of DCE as

$$\varsigma \sim \frac{\hbar \omega_0^2 l_{SPP}}{\lambda_0^3} \eta_0^2 \eta^4 \left|\chi^{(2)}\right|^2 Z_0 \sim 10^{-6}. \tag{26}$$

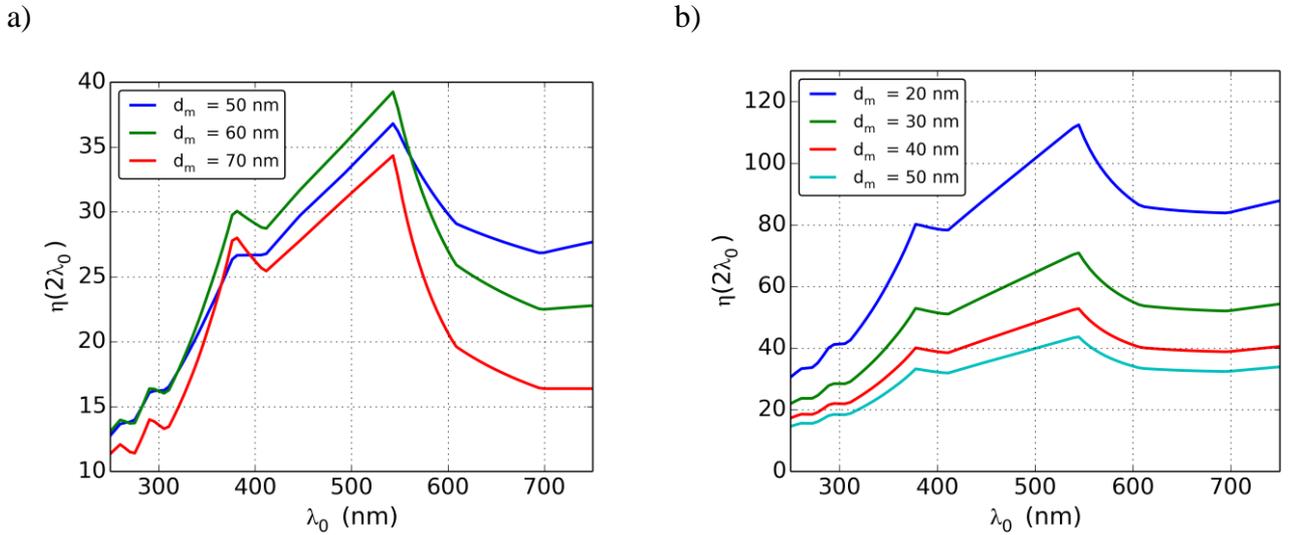

Fig. 4. The enhancement of the electric field of SPPs at air-silver interface (at the wavelength twice the wavelength of laser light $\lambda_0$). Fig. (a) and (b) corresponds to different setups in Fig. 1a and Fig. 1b respectively. The refractive index of prism $n_0 = 1.5$ and the dielectric is taken to be air, e.g. $n_1 = 1.0$.

## 7. Concluding remarks

A theoretical consideration of the dynamical Casimir effect in a metal-dielectric interface with asymmetric dielectric is presented. It is shown that the enhancement of the field of surface plasmon polaritons in the interface may allow one to generate photon pairs with remarkable yield of the order of $10^{-6}$ or more. The efficiency of the dynamical Casimir effect in a metal-dielectric interface could be additionally enhanced a few orders of magnitude if to use a proper grating: in addition to the enhancement of the field of generated surface plasmon polaritons this would allow one to enhance the field of excitation as well. One more possibility to enhance the efficiency of the process under consideration is to surround the metallic film by layers of dielectric crystals with enhanced second order susceptibility. As examples of such crystals may serve AgGaSe2 and NiNbO3, where $\chi^{(2)} \gtrsim 30$ pm/V [26]. It is expected that analogous value of $\chi^{(2)}$ should have chalcopyrite compounds [27]. Moreover hetero-structures with asymmetric quantum wells presumably may have $\chi^{(2)} > 400$ pm/V [28]. This may allow one to additionally increase the yield $\varsigma$ of the process a few orders of magnitude. It is not excluded that the dynamical Casimir effect in metal-dielectric interfaces may allow one to achieve full conversion of incident photons to photon pairs.

**Acknowledgements**
The research was supported by Estonian research projects SF0180013s07, IUT2-27 and by the European Union through the European Regional Development Fund (project 3.2.0101.11-0029).

**References**
[1] D. C. Burnham and D. L. Weinberg, "Observation of simultaneity in parametric production of optical photon pairs", *Phys. Rev. Lett.* **25**, 84 (1970).
[2] B. E. Saleh and M. C. Teich, "Fundamentals of Photonics", *Wiley Series in Pure and Applied Optics, Wiley-Interscience* (2007).
[3] G.T. More, "Quantum Theory of the Electromagnetic Field in a Variable Length One Dimensional Cavity ", *J. Math. Phys* **11,** 2679 (1970).
[4] A. Lobashev, V.V. Mostepanenko, "Quantum effects in nonlinear insulating materials in the presence of a nonstationary electromagnetic field", *Theor. Math. Phys.* **86**, 3 (1991).
[5] V.Hizhnyakov, "Quantum emission of a medium with a time-dependent refractive-index", *Quantum Opt.* **4**, 277 (1992).
[6] V.V. Dodonov, A.B. Klimov, D.E. Nikonov, "Quantum phenomena in nonstationary media", *Phys. Rev. A* **47**, 4422 (1993).
[7] H. Johnston, S. Sarkar, "Moving mirrors and time-varying dielectrics", *Phys. Rev. A* **51**, 4109 (1995).
[8] M. Cirone, K. Rzazewski, J. Mostowski, "Photon generation by time-dependent dielectric: A soluble model" *Phys. Rev. A* **55**, 62 (1997).
[9] V. Hizhnyakov, "Strong two-photon emission by a medium with periodically time-dependent refractive index", *arXiv:quant-ph/0306095v1* (2013).
[10] V. Hizhnyakov, H. Kaasik, "Dynamical Casimir effect: quantum emission of a medium with time-dependent refractive index", *IEEE Conference Proceedings: Northern Optics 2006* (2006).

[11] H. Saito, H. Hyuga, "The dynamical Casimir effect for an oscillating dielectric model", *J. Phys. Soc. Jpn.* **65**, 3513 (1996).

[12] M. Bordag (editor), "The Casimir Effect 50 Years Later", *Proceedings of the Fourth Workshop on Quantum Field Theory under the Influence of External Conditions, World Scientific, Leipzig* (1999).

[13] C.M. Wilson, G. Johansson, A. Poukabirian, M. Simoen, J.R. Johansson, T. Duty, F. Nori and P. Delsing, "Observation of the dynamical Casimir effect in a superconducting circuit", *Nature* **376**, 479 (2011).

[14] V. Hizhnyakov, H. Kaasik, I. Tehver, "Spontaneous nonparametric down-conversion of light" *Applied Physics A. Materials Science & Processing: META'13 4th International Conference on Metamaterials, Photonic Crystals, and Plasmonics; Sharjah, United Arab Emirates, Springer Verlag*, 563 – 568 (2014).

[15] S. Maier, "Plasmonics: Fundamentals and Applications", *Springer* (2004).

[16] L. Novotny and B. Hecht, "Principles of Nano-Optics", *Cambridge University Press* (2006).

[17] D. Sarid, "Long-range surface-plasma waves on very thin metal films", *Physical Review Letters* **47**, 1927 (1981).

[18] J.J. Burke, G.I. Stegman, T. Tamir, "Surface-polariton-like waves guided by thin, lossy metal films", *Phys. Rev. B* **33**, 5186 (1986).

[19] Z.H. Han, S.I. Bozhevolnyi, "Radiation guiding with surface plasmon polaritons", *Reports on Progress in Physics* **76** (2013).

[20] P.B Johnson and R. W. Christy, "Optical constants of the noble metals", *Phys. Rev. B* **6**, 4370 (1972).

[21] N.D. Birrell, P.C.W. Davies, "Quantum field in curved space", *Cambridge University Press* (1982).

[22] A. Ling, A. Lamas-Linares and C. Kurtsiefer, "Absolute emission rates of spontaneous parametric down-conversion into single transverse Gaussian modes", *Phys. Rev. A* **77**, 043834 (2008).

[23] V. Hizhnyakov, A. Loot, "Spontaneous down conversion in metal-dielectric interface - a possible source of polarization-entangled photons", *arXiv:1406.2174 [quant-ph]* (2014).

[24] F. Abeles, "Optical Properties of Thin Absorbing Films", *J. Opt. Soc. Am.* **47**, 473 (1957).

[25] J. Chilwell and I. Hodgkinson, "Thin-films field-transfer matrix theory of planar multilayer waveguides and reflection from prism-loaded waveguides", *J. Opt. Soc. Am.* **1**, 7 (1984).

[26] R.W. Boyd and G.L. Fischer, "Encyclopedia of Materials: Science and Technology", ISBN:0-08-0431526, pp. 6237-6244 (2011).

[27] A. H. Reshak, M. G. Brik, and S. Auluck, "Dispersion of the linear and nonlinear optical susceptibilities", *Journal of Aplied Physics* **116**, 103501 (2014).

[28] A. Rostamia, H. Baghbana, H. Rasooli Saghaib, "An ultra-high level second-order nonlinear optical susceptibility in strained asymmetric GaN–AlGaN–AlN quantum wells: Towards all-optical devices and systems", *Microelectronics Journal* **38**, 900 (2007).